\documentclass[conference]{IEEEtran}
\IEEEoverridecommandlockouts
\usepackage{cite}
\usepackage{amsmath,amssymb,amsfonts}
\usepackage{algorithmic}
\usepackage{graphicx}
\usepackage{textcomp}
\usepackage{xcolor}
\usepackage{bm}
\def\BibTeX{{\rm B\kern-.05em{\sc i\kern-.025em b}\kern-.08em
    T\kern-.1667em\lower.7ex\hbox{E}\kern-.125emX}}

\usepackage{url}
\usepackage{paralist}

\newcommand{\pmtwo}{PM\textsubscript{2.5}}
\newcommand{\pmten}{PM\textsubscript{10}}
\newcommand{\sotwo}{SO\textsubscript{2}}
\newcommand{\notwo}{NO\textsubscript{2}}

\newcommand{\name}{AQEyes}
\newcommand{\cc}[1]{\textcircled{\small{#1}}}
\newcommand{\ano}[1]{$\textbf{#1}$}

\begin{document}

\title{\name: Visual Analytics for Anomaly Detection and Examination of Air Quality Data}


 \author{\IEEEauthorblockN{Dongyu Liu}
 \IEEEauthorblockA{\textit{CSE} \\
 \textit{HKUST}\\
 Hong Kong, China \\
 dliuae@cse.ust.hk}
 \and

 \IEEEauthorblockN{Kalyan Veeramachaneni}
 \IEEEauthorblockA{\textit{LIDS} \\
 \textit{MIT}\\
 Cambridge, MA, USA \\
 kalyan@csail.mit.edu}
 \and

 \IEEEauthorblockN{Alexander Geiger}
 \IEEEauthorblockA{\textit{LIDS} \\
 \textit{MIT}\\
 Cambridge, MA, USA \\
 geigera@mit.edu}
 \and

 \IEEEauthorblockN{Victor O.K. Li}
 \IEEEauthorblockA{\textit{EEE} \\
 	\textit{HKU}\\
 	Hong Kong, China \\
 	vli@eee.hku.hk}
 \and

 \IEEEauthorblockN{Huamin Qu}
 \IEEEauthorblockA{\textit{CSE} \\
 	\textit{HKUST}\\
 	Hong Kong, China \\
 	huamin@cse.ust.hk}
 }


\maketitle

\begin{abstract}

Anomaly detection plays a key role in air quality analysis by enhancing situational awareness and alerting users to potential hazards. However, existing anomaly detection approaches for air quality analysis have their own limitations regarding parameter selection (e.g., need for extensive domain knowledge), computational expense, general applicability (e.g., require labeled data), interpretability, and the efficiency of analysis. Furthermore, the poor quality of collected air quality data (inconsistently formatted and sometimes missing) also increases the difficulty of analysis substantially. In this paper, we systematically formulate design requirements for a system that can solve these limitations and then propose \name, an integrated visual analytics system for efficiently monitoring, detecting, and examining anomalies in air quality data. In particular, we propose a unified end-to-end tunable machine learning pipeline that includes several data pre-processors and featurizers to deal with data quality issues. The pipeline integrates an efficient unsupervised anomaly detection method that works without the use of labeled data and overcomes the limitations of existing approaches. Further, we develop an interactive visualization system to visualize the outputs from the pipeline. The system incorporates a set of novel visualization and interaction designs, allowing analysts to visually examine air quality dynamics and anomalous events in multiple scales and from multiple facets. We demonstrate the performance of this pipeline through a quantitative evaluation and show the effectiveness of the visualization system using qualitative case studies on real-world datasets.

\end{abstract}



\section{Introduction}\label{sec:introduction}

The rapid processes of industrialization and urbanization have greatly improved the economy while also intensifying air pollution issues, a condition which causes tremendous physical and psychological harms to humans.
A report from WHO has shown that ambient air pollution caused around 4.2 million premature deaths worldwide in 2016\footnote{\url{https://www.who.int/en/news-room/fact-sheets/detail/ambient-(outdoor)-air-quality-and-health}}.
Thus, monitoring the dynamics of air pollutants is an important and pressing task.
Failure to detect and respond to unusual changes (e.g., an outbreak) of air pollutants will both create enormous risks to human health and cause great loss to our economy.
Anomaly detection, therefore, becomes an essential part of air quality analysis. 
Detecting anomalies is useful in quickly identifying an air pollution \textit{event} that is defined as a valid observation of unexpected air pollutant concentrations at a certain time period and place compared with previous observations from that location~\cite{zhang2007taxonomy}.
Based on experience, air quality events can result from unusual weather conditions or some local sources such as trash burning.

Currently, anomaly detection for air quality data primarily utilizes statistical and threshold-based~\cite{paschalidis2010statistical}, density-based~\cite{chandola2009anomaly}, learning-based (need labeled data)~\cite{ayadi2015machine, hill2007real}, and visualization-based approaches~\cite{lee2009visualization, dang2012timeseer}.
Each of these approaches has drawbacks. 
Statistical and threshold-based methods require extensive human knowledge to specify model parameters and become problematic when parametric assumptions are violated.
Density-based methods are computationally intensive, and it is challenging to define the distance between multivariate measurement data.
Both statistics-based and density-based methods are unable to capture anomalies that are characterized by temporal trends.
Moreover, learning-based methods require high-quality labeled data that is often unavailable or too time-consuming to collect.
Visualization-based methods allow to flexibly and adaptively identify and interpret anomalies, but they also require analysts to manually observe multiple air quality variables, which becomes more impractical as the amount of data increases.
Hence, to ensure efficiency, accuracy, and interpretability, a system that integrates a more accurate, scalable, and intelligent anomaly detection method with application-tailored visualization techniques is needed.

The following four key technical challenges for designing that system can be identified.
First, air quality analysis usually involves pollutant and weather data collected in multiple air quality monitoring stations and in different time granularities, leading to inconsistently-formatted data.
The data is also often missing due to unavoidable factors (e.g., malfunction of sensors).
Hurdles like these increase the difficulty of obtaining high-quality inputs for the use of anomaly detection.
Second, the lack of labeled anomalies necessitates the use of unsupervised or semi-supervised approaches.
Third, a single model is insufficient to handle all situations, because stations in different regions have different environments and standards concerning anomalies.
Lastly, the large scale of the data introduces many obstacles in designing a visualization system to support efficient anomaly pattern exploration and examination.

In this paper, we systematically formulate the system design requirements and then propose \name, an integrated visual analytics system for efficiently monitoring, detecting, and examining anomalies in air quality data.
The major contributions of our work are summarized as follows:

\begin{itemize}
    \item We propose a unified end-to-end tunable machine learning pipeline, which solves the problems of missing and differently-granularized data and integrates an efficient unsupervised anomaly detection method adapted and extended from other domains.
    
    \item We propose several novel visualization and interaction designs to cooperate seamlessly with the machine learning pipeline. The designs enable analysts to efficiently explore and examine air quality dynamics and anomalous events from different perspectives and levels of detail.
    
    \item We evaluate the machine learning pipeline and visualization designs through both quantitative and qualitative case studies on two real-world air quality data sets.
  
\end{itemize}



\section{Related Work}\label{sec:relatedwork}

\subsection{Air Quality Analysis}

For many years, researchers from various domains have spent a great deal of effort on the development of data analysis techniques for understanding air quality.
Several excellent surveys summarize the techniques well~\cite{hodge2004survey, chandola2009anomaly, zheng2014urban, rautenhaus2018visualization}.
In the following we will focus on the most relevant work.

\subsubsection{\textbf{Anomaly detection is a key task in air quality analysis}}
The general approach of anomaly detection is to find unexpected patterns in data.
The simplest anomaly detection approaches are out-of-limits methods which flag locations surpassing predefined thresholds on raw values.
A number of other more complex anomaly detection techniques have been proposed as improvements on out-of-limits approaches.
These can be divided into four categories, including statistics-based ~\cite{paschalidis2010statistical, shaadan2015anomaly, sguera2016functional}, density-based~\cite{knorr2000distance, bay2003mining, breunig2000lof, zhang2010outlier}, learning-based (usually requiring labeled data)~\cite{ayadi2015machine, hill2007real}, and visualization-based approaches~\cite{lee2009visualization, dang2012timeseer}.
Though these methods have demonstrated their effectiveness in a variety of  scenarios~\cite{chandola2009anomaly}, each of them has limitations related to parameter selection (e.g., need for domain knowledge), computational expense, general applicability (e.g., require labeled data), interpretability, and the efficiency of analysis (see Section~\ref{sec:introduction}).

In recent years, recurrent neural networks (RNNs) have achieved huge success, leading to a performance breakthrough in sequence-to-sequence learning tasks~\cite{sutskever2014sequence}.
Long Short-Term Memories (LSTMs), a special type of RNNs, have proved good performance in learning the relationship between past and current data values.
LSTMs can handle multivariate time-series without the strong need for application domain knowledge.
Hence, LSTMs are widely used in time series forecasting and anomaly detection~\cite{malhotra2015long, bontemps2016collective, malhotra2016lstm, hundman2018detecting}.
The basic idea is to fit LSTM models on normal time series data and compare model predictions to actual data values with a set of anomaly detection rules.
However, simply applying these approaches on air quality data analysis is difficult due to the technical challenges introduced in Section~\ref{sec:introduction}.
We thereby introduce a modular end-to-end machine learning pipeline that incorporates various preprocessing steps, an LSTM model and a dynamic error processing, to detect anomalous sequences in the time series in an unsupervised manner.
The advantage of the pipeline is the simplicity by which the single module of the pipeline can be changed, allowing analysts to easily develop multiple pipelines and evaluate their performances.



\subsubsection{\textbf{Visual analytics is an important tool for air quality analysis}}
Visualization exploits humans excellent ability to perceive visual patterns, thereby seamlessly connecting humans to the data analysis process.
Visualizations incorporating appropriate data reduction techniques can provide analysts with a straightforward and natural way to monitor multiple air quality variables and their evolutions~\cite{lee2009visualization}.
Qu et al.~\cite{qu2007visual} present a comprehensive system to analyze the air pollution problem in Hong Kong, where a series of novel visualizations such as circular bar charts and weighted complete graphs, are integrated to investigate the correlation between multiple attributes.
Chen et al.~\cite{guo2019visual} introduce a novel tree structure to organize the correlations among air quality variables, enabling analysts to monitor the evolving correlations among these variables.
Quinan and Meyer~\cite{quinan2015visually} propose a set of encoding choices and interaction methods to interpret how multiple weather features relate to forecasting outcomes for more precise results.
Du et al.~\cite{du2017visual} develop an interactive visualization system to support efficient exploration of air quality data at multiple scales.

To the best of our knowledge, our system is the first comprehensive visual analytics system that is primarily designed for anomaly detection, exploration and assessment of air quality data.
The system is built on a machine learning pipeline, which can not only produce comprehensible results more efficiently but also provide various ways for analytsts to interact with the results in rich spatiotemporal context.

\subsection{Multivariate Spatial Time-series Visualization}
Air quality data can be regarded as multivariate time-series in spatial context.
Each weather variable or pollutant is a time-varying attribute associated with a stationary location (i.e., an air quality monitoring station).
Hence, the techniques used for visualizing and structuring time-series can also be applied in air quality analysis.
The key difference lies in the way the timeline is encoded.

Time is linear but contains an inherent hierarchical structure of granularities, such as hours, days, weeks, and months.
A standard method to visualize a time-series is mapping time to the horizontal x-axis and time-dependent variables to the vertical y-axis~\cite{aigner2011visualization}.
When we want to observe cyclic/periodical patterns, a spiral-shaped time axis ~\cite{weber2001visualizing} is a useful time-encoding scheme.
If the timeline emphasizes individual dates, a calendar layout to represent time would be more suitable to depict the daily, monthly, or yearly value changes~\cite{van1999cluster}.

To further add the spatial dimensions, one popular solution is providing several separated views to display information regarding space or time.
These views are then linked together via user interactions for coordinated analysis~\cite{liu2017smartadp}.
Other approaches attempt to simultaneously encode spatial, temporal, and other attributes in one view, such as ring maps~\cite{zhao2008activities}, glyphs on maps~\cite{andrienko2017revealing}, a space-time cube~\cite{kraak2003space}, and small multiples~\cite{beecham2017map}.


Inspired by the early design guidelines and visualization designs, in this work, we propose a set of novel hybrid visualization and interaction designs.
These designs are particularly suitable for interpreting the results returned by the machine learning pipeline.


\section{Problem Formulation}\label{sec:background}


\subsection{Data Abstraction}\label{sec:background:data}
We mainly use two air quality datasets which cover three different big cities in China.
Dataset A covers two cities (Beijing and Shenzhen) with 50 air quality monitoring stations (a data subset from the work~\cite{zheng2015forecasting, zheng2013u}).
Millions of air quality and weather records are collected over a period of one year (from 2014/05/01 to 2015/04/30).
Each station documents attributes including 6 types of pollutants including Carbon monoxide (CO), Nitrogen dioxide (\notwo), Ozone (O\textsubscript{3}), Sulphur dioxide (\sotwo), \pmtwo, and \pmten, and 4 types of weather variables including temperature, pressure, humidity, and wind speed.
Each attribute represents a sequence of time-varying values (i.e., a time-series).
Dataset B includes 16 air quality monitoring stations in Hong Kong recording air quality and weather information from 2016 to present.
The data is provided by the Environmental Protection Department and the Hong Kong Observatory of the Hong Kong Special Administrative Region.
%

\subsection{Requirement Analysis}

The primary goal of this work is to help domain experts explore air quality dynamics and identify anomalous events for further examination.
Over the past six months, we have worked with two domain experts that have considerable experience in air quality data analysis.
After a series of conversations, we compiled the following system design requirements.

\begin{enumerate}[R.1]
	\item \textbf{Handle data at different time granularities and data missing issues}.
		  Different data attributes at different monitoring stations could be collected at different time granularities.
		  For example, station A collects \notwo~information every hour and the temperature information every 3 hours, but station B collects \notwo~information every 2 hours and temperature information every half hour.
		  Also, data collection could be missed due to sensor malfunction or maintenance.
		  Both issues increase the difficulty of data processing and analyzing substantially.
		  \textit{Can we have a unified data process and analysis pipeline to overcome these issues effectively?}

    \item \textbf{Support effective and smart anomaly identification}.
		  Thresholds on raw values are often based on statistical information and domain knowledge to enable the detection of anomalies regarding air pollutants.
		  However, conditions at different stations differ from each other and evolve over time.
		  Experts have to continually adjust these thresholds to cope with both of these problems.
		  Moreover, threshold-based methods are unable to detect contextual anomalies~\cite{ahmad2017unsupervised}, which are unexpected events that may happen within thresholds but significantly disobey temporal trends.
		  \textit{Can we efficiently detect anomalous sequences in the data set, while not relying on fixed thresholds, and reduce the number of false positives while still flagging anomalous events?}
	
	\item \textbf{Provide efficient visual designs to explore and assess anomalous events.}
		   One significant concern is the overwhelming amount of available data.
		   One data attribute (e.g., \pmtwo) at one station refers to one time series.
		   Thus, thousands of time series need to be monitored at any given time.
		   Analysts are only able to look at a small subset of the data at the same time, which means a large number of adverse events could be missed.
		   \textit{What kinds of visualization techniques and designs can be used to facilitate the efficient exploration and assessment of anomalous events?}
		   
		   
    \item \textbf{Enable multi-scale and multi-facet visual exploration of air quality dynamics}.
		  The volume of data does not allow us to show every detail.
		  Thus, the visualization should follow the mantra ``overview first, zoom and filter, then details-on-demand''~\cite{shneiderman2003eyes} to display the data in different levels of details.
		  Moreover, the time dimension itself is extremely complex, containing multiple levels of hierarchical structures (e.g., day, week, month, and year) and natural cycles (events happening periodically).
		  Hence, the system should also provide views to explore air quality dynamics from a different perspective.		   

	\item \textbf{Allow comparative analysis between attributes and between stations.}
		  Different pollutants could have different temporal patterns.
		  Also, air quality dynamics in different regions could be very different.
		  Observing the similarities and differences between attributes and between regions would offer significant insights.
		  
	\item \textbf{Support interactive labeling and feedback gathering.}
		  It cannot be guaranteed that automatic methods will identify all anomalous events correctly.
		  Therefore, analysts should be allowed to flexibly add, modify, or delete an event, or to make comments or tags on events.
		  These comments are not only helpful in recording the analysis process but also useful in adapting existing models over time for better anomaly identification in the future.

\end{enumerate}


\section{system}\label{sec:system}

\begin{figure*}[!htbp]
	\centering
	\vspace{-8pt}
	\includegraphics[width=0.8\linewidth]{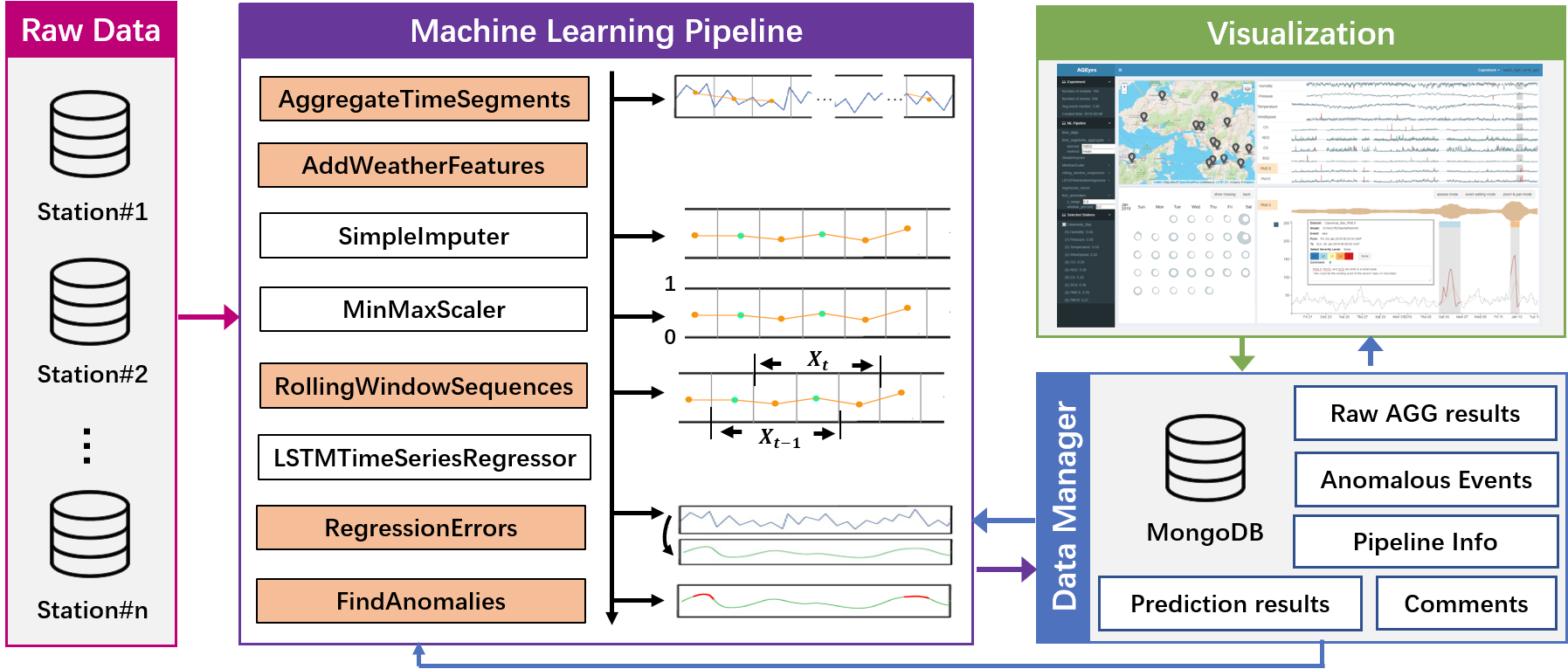}
	\caption{The system architecture of \name.}
	\vspace{-10pt}
	\label{fig:system_overview}
\end{figure*}

Fig.~\ref{fig:system_overview} presents the system architecture that comprises three major modules (b, c, d), namely, machine learning pipeline (ML pipeline), data manager, and visualization.
The raw data of one pollutant at a monitoring station is fed into the ML pipeline for processing (\textbf{R1, R2}).
The computing results of the pipeline, such as the time-series after aggregation, LSTM's prediction results, and identified anomalous events, are collected by the data manager.
The collected data is then used in a visualization module (\textbf{R3, R4, R5}), where analysts can explore the rich dynamics of air pollutants and weather variables, interact with the anomalous events (\textbf{R6}), and iteratively tune the hyperparameters of the pipeline.

\subsection{Machine Learning Pipeline for Air Quality Analysis}

To tackle the different data granularities and missing data issues and support efficient anomaly identification (\textbf{R1, R2}), 
we build a unified end-to-end tunable machine learning pipeline particularly for air quality analysis by using the MLBlocks framework~\cite{xue2018flexible}.
The pipeline consists of many different components (so called blocks), each of which has its own task. The blocks are reusable components and can be stacked in an arbitrary order to create pipelines. Most of the blocks belong to one of the following two classes: ``transform`` and ``learning``.
``Transform blocks`` simply receive data, perform computations, and return the resulting data.
``Learning blocks`` are able to learn parameters of a function before applying the function to the data and returning the output. These blocks can be any classifier or regressor. 

As shown in Fig.~\ref{fig:system_overview}, the pipeline we are proposing is composed of a set of blocks: multiple ``transform blocks`` that are used for preprocessing and anomaly detection, and a ``learning block`` that implements the trainable LSTM model.
The blocks in orange windows are customized for air quality analysis, whereas the remaining blocks are more general blocks.
When the raw time-series data of a certain pollutant at one station is input into the pipeline,
the first block performs segmentation operation and produces a sequence taken at successive equally spaced points in time.
The second block adds weather information to the original sequence and outputs a multi-dimensional feature vectors shown in Fig.~\ref{fig:model_input}.
The third block is then applied for imputing missing values.
The current feature vectors are then normalized and transformed for the use of training an LSTM model.
After that, the prediction errors are computed, smoothed, and passed to the last block for anomaly identification.

There are two significant advantages to building such a pipeline using MLBlocks.
First, the pipeline specifies and exposes hyperparameters in a clean and simple manner.
This allows us to provide analysts with an interactive way to tune the hyperparameters for rapid experimental verification.
Second, the pipeline is general and flexible enough, where each block is reusable and replaceable.
It is easy to adapt the pipeline to other air quality analysis tasks (such as pollutant correlation analysis), or even common spatio-temporal data analysis tasks, with a minimum of effort.

In the following, we will introduce the LSTM and anomaly detection blocks in detail, which are the two core blocks in our proposed pipeline.

\subsubsection{\textbf{Pollutant value prediction with LSTMs}}
The block ``LSTMTimeSeriesRegressor'' aims to create an LSTM model for a monitoring station to predict values for a certain pollutant.
There are several reasons why we model each pollutant independently.
First, it is common to treat different pollutants separately, as different air pollutants are influenced by different weather factors in varying degrees~\cite{zheng2013u}.
Second, a model for one pollutant allows traceability down to a specific pollutant level, making the anomalies detected then easier to interpret.

The intuition behind the use of LSTMs in time-series anomaly detection is the fact that LSTMs are able to learn both very complex representations of input data and relations between the single steps in the input sequence.
This is very useful when trying to accurately model time-series data.

Consider a time-series $X = \{\textbf{x}^1, \textbf{x}^2, \cdots, \textbf{x}^n\}$ where $\textbf{x}^t$ is a 5-dimensional vector $\{w^t_1, w^t_2, w^t_3, w^t_4, x^t\}$.
In the vector, $x^t$ refers to the monitored value of a certain pollutant at a given station at time step $t$,
and $w^t_1$, $w^t_2$, $w^t_3$, $w^t_4$ denote the weather variables regarding temperature, humidity, barometric pressure, and wind speed, respectively.
The LSTM takes the sequence as input and computes the output by applying several computations to each of the inputs in the sequence, always taking into account also the output of the previous input.
In our case, the output of the LSTM is input to another dense layer, which outputs one scalar as the prediction. 

\begin{figure}[!tbp]
	\centering
	\includegraphics[width=1\linewidth]{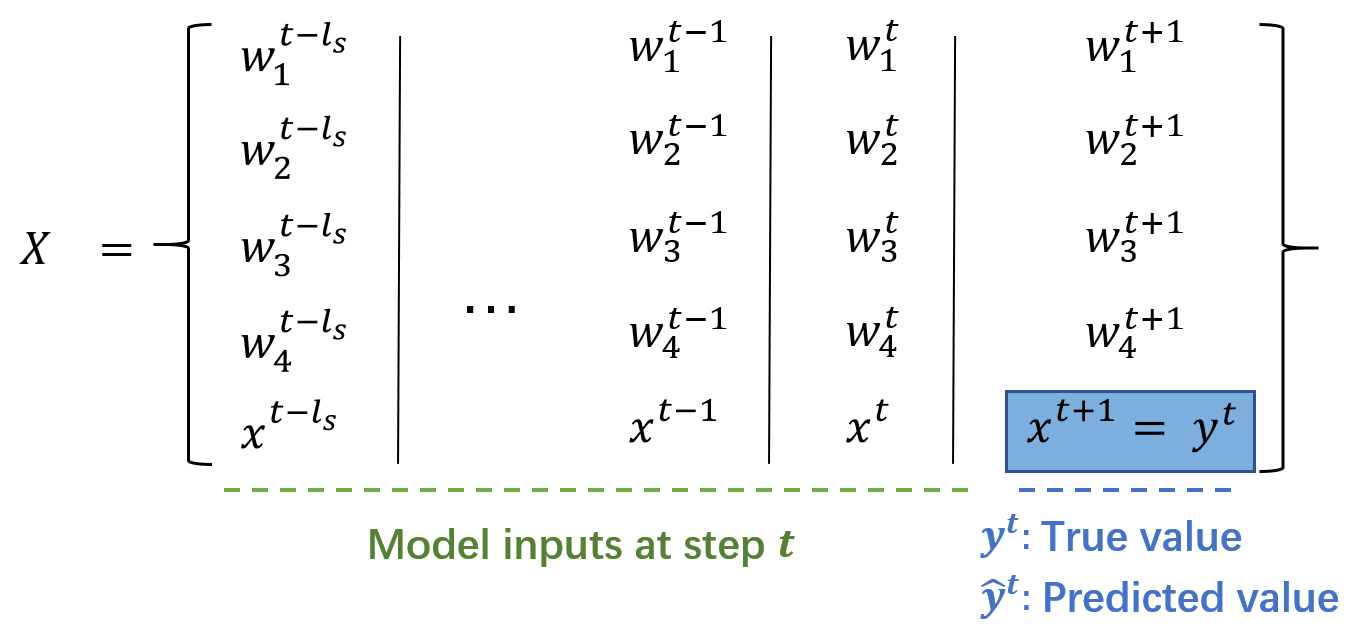}
	\caption{An illustration of the input matrices used for prediction at each time step $t$. The prediction errors $e^t = |\hat{y}^t - y^t|$ will be used to determine whether time step $t$ is anomalous.}
	\vspace{-16pt}
	\label{fig:model_input}
\end{figure}

At time step $t$, the model is fed with a sequence of continuous $\textbf{x}^i$ with length $l_s$ in order to predict the subsequent pollutant value $\hat{y}_{t}$ (as shown in Fig~\ref{fig:model_input}). Then, the prediction error at time step $t$ can be computed as $e^t = |\hat{y}^t - y^t|$, where $y^t = x^{t+1}$.

\subsubsection{\textbf{Anomaly detection with dynamic error thresholds}}

The intuition is that exceptionally high prediction errors could suggest some unexpected behaviors.
Thus, we borrow an unsupervised method by Hundman et al.~\cite{hundman2018detecting}, where an error threshold can be efficiently learned from error sequences without labeled data or statistical assumptions about errors.
To identify whether a pollutant behaves abnormally at time step t, an error sequence $\bm{e} = [e^{t-h}, \cdots, e^{t-1}, e^{t}]$ is evaluated, where $h$ specify the number of historical errors.
To better capture the continuous anomalous time steps, we employ a simple moving average (ma) over the errors to get a smoothed set of errors:
$$\bm{e}_s = \text{ma}(\bm{e}) = [e^{t-h}_s, \cdots, e^{t-1}_s, e^{t}_s]$$
The threshold is then selected from the set: $$\bm{\theta} = \mu(\bm{e}_s) + k\sigma(\bm{e}_s)$$
where $\mu$ and $\sigma$ denote the mean and standard deviation respectively and $k \in \mathbb{R}_0^+$  (one hyperparameter of this block and in practice, we often consider $k \in [0, 12]$).
Eventually, $\theta$ is determined by:
$$\underset{\bm{\theta}}{argmax} \quad \frac{\Delta\mu(\bm{e}_s)/\mu(\bm{e}_s) + \Delta\sigma(\bm{e}_s)/\sigma(\bm{e}_s)}{|\bm{e}_a| + |\bm{E}_{seq}|^2}	$$
where
\begin{align*} 
&\Delta\mu(\bm{e}_s) = \mu(\bm{e}_s)-\mu(\{e_s \in \bm{e}_s, | e_s < \theta\}) \\
&\Delta\sigma(\bm{e}_s) = \sigma(\bm{e}_s)-\sigma(\{e_s \in \bm{e}_s, | e_s < \theta\}) \\
&\bm{e}_a = \{e_s \in \bm{e}_s | e_s > \theta\} \\
&\bm{E}_{seq} = \text{continuous sequences of } e_a \in \bm{e}_a
\end{align*}

The intuition is that finding a threshold that would bring about the maximum percent decrease in the mean and standard deviation of $\bm{e}_s$ if all error values above the threshold are eliminated.
$\bm{E}_{seq}$ and $\bm{e}_a$ in the denominator are introduced as penalization terms to prevent having too many anomalous error values and sequences by simply choosing a very small threshold.
Now each anomalous sequence $\bm{e}_{seq} \in \bm{E}_{seq}$ can be assigned with an anomaly score $s$:
\begin{equation}
s = \frac{max(\bm{e}_{seq}) - \theta}{\mu(\bm{e}_s) + \sigma(\bm{e}_s)}
\label{eq:score}
\end{equation}

\subsection{Visualization and Interaction}

\begin{figure*}[!htbp]
	\centering
	\vspace{-18pt}
	\includegraphics[width=1\linewidth]{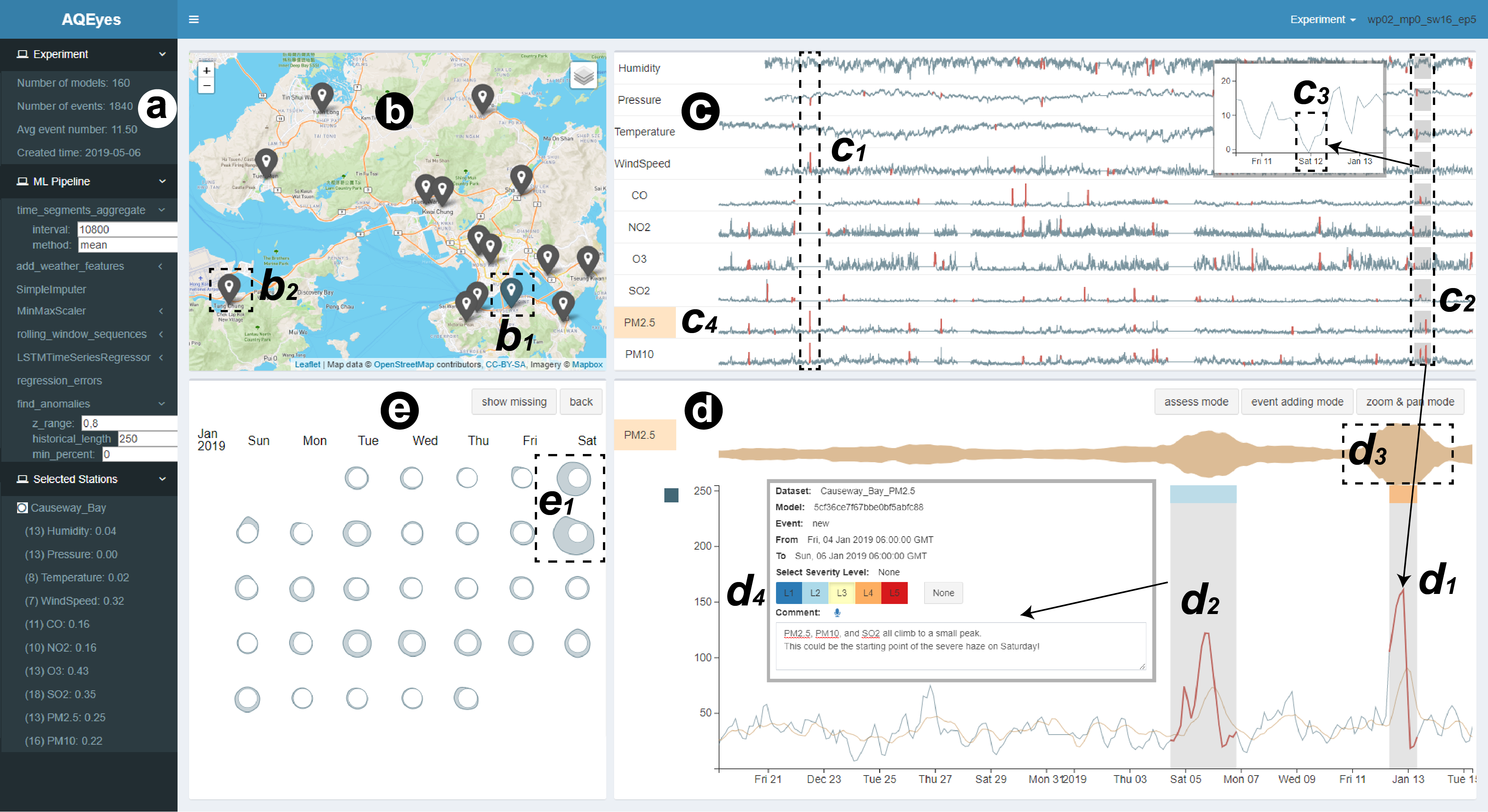}
	\vspace{-16pt}
	\caption{The user interface of \name. \cc{a} Dashboard View shows the information about the current experiment, used ML pipeline, and selected stations. \cc{b} Map View allows analysts to choose one or multiple stations for examination. \cc{c} Time-series Context view presents an overview of the dynamics of each pollutant and weather variable of the selected stations. \cc{d} Time-series Focus view details the context view, providing various information to evaluating the pipeline and the identified anomalies. \cc{e} Period View integrates periodical glyphs in three levels (year, month, and day) to help reveal the periodic patterns.}
	\vspace{-16pt}
	\label{fig:interface}
\end{figure*}

This section describes the key visualization and interaction techniques that assist analysts in exploring air quality dynamics and assessing anomalous events (\textbf{R3, R4, R5, R6}).
Fig.~\ref{fig:interface} presents the overview of our interface, which is composed of five main sub-views, namely: dashboard view \cc{a}, map view \cc{b}, time-series context view \cc{c}, time-series focus view \cc{d}, and period view \cc{e}.
We define an \textit{experiment} as a process where analysts have tried some hyperparameters (e.g., LSTM training epoch) of the ML pipeline and then run the pipeline on a set of stations, producing a number of LSTM models and anomalous events.
The multiple views work together for a coordinated analysis of an experiment which can be selected at the top-right of the interface.

\subsubsection{\textbf{Dashboard View}}
The dashboard view (Fig.~\ref{fig:interface}\cc{a}) lists information that is helpful for analysts to roughly evaluate whether an experiment has achieved their expectation.
From top to bottom, it first shows the summary information of the currently selected experiment, including the total number of models (each pollutant at one station has one model) and anomalous events, the average number of events detected by one model, and experiment creating time.
Then, the pipeline structure together with the user-specified hyperparameters of each block are presented.
Analysts can edit these parameters and save them into database for the next experiments.
The bottom of the dashboard displays the information regarding the stations selected in the map view.
For example, ``\mbox{(10) \notwo: 0.16}'' can be interpreted as that 10 anomalous events regarding \notwo~are identified at the station Causeway Bay and the Mean Absolute Percentage Error (a measure of prediction accuracy of a forecasting method) of the LSTM model is 0.16.

\subsubsection{\textbf{Map View}}
One special aspect of air quality data is that the time-series data of a pollutant is highly associated with a spatial location (i.e., air quality monitoring station).
Therefore, we provide a map view (Fig.~\ref{fig:interface}\cc{b}) to offer analysts a map-centered exploratory approach so that they can obtain the most intuitive insights into both the environment and the spatial relationships between stations.
Each monitoring station is visualized with a grey map marker.
Analysts can click on a marker to select a station for further observation (\ano{b1} in Fig.~\ref{fig:interface}\cc{b}).
Multiple stations can be selected at the same time for a comparative analysis (\textbf{R5}).
The selected stations are highlighted with categorical colors consistent with the colors used in other views to display station information (Fig.~\ref{fig:hk_compare}).

\subsubsection{\textbf{Time-series Context View}}
The time-series context view (Fig.~\ref{fig:interface}\cc{c}) presents analysts an overview to quickly scan through the dynamics of each pollutant and weather variable (\textbf{R4, R5}).
Once a station is selected on the map, the multiple time-series associated with the station are plotted as a juxtaposition (i.e, small multiples); the top 4 rows are the weather variables and the bottom 6 rows are the 6 types of pollutants.
The timeline of each time-series is aligned horizontally and the identified anomalous events   are highlighted with red curve segments (\textbf{R3}).
This thus facilitates the comparative analysis among multiple attributes of a station. 
The anomalous events from different time-series occurring in the same time segment could suggest a rare event that has significant impacts on air quality (\ano{c1} and \ano{c2} in Fig.~\ref{fig:interface}\cc{c}). 

The first design consideration is the problem of displaying all time-series given limited screen space. Thus, a highly space-efficient design should be considered.
Given that different time-series have distinct value units and scales, plotting all of them in one graph is unreasonable.
Hence, we adopt small multiples~\cite{javed2010graphical}, one of the most popular space-efficient techniques for visualizing multiple time-series.
The second consideration is which kinds of charts (e.g., line chart, area chart, horizon graph, and box plots) are more effective in this application scenario.
After several rounds of trials and discussion with the experts, we finally adopt line charts, as line charts are not only easier for individual value examination but also good for trend tracking~\cite{javed2010graphical, munzner2014visualization}; both features are particularly suitable for the scenarios involving anomaly analysis.
More importantly, when multiple stations are selected, the time-series of a certain pollution or weather variable from different stations can be easily plotted in one chart for comparative analysis (Fig.~\ref{fig:hk_compare}\cc{a} and Fig.~\ref{fig:bjsz}\cc{a}).

\subsubsection{\textbf{Time-series Focus View}}
Analysts can select one time-series from the context view (\ano{c4} in Fig.~\ref{fig:interface}\cc{c}) for a detailed analysis in the focus view (Fig.~\ref{fig:interface}\cc{d}, \textbf{R4}).
The focus view extends a piece of the timeline (the time segments with grey background, \ano{c2}) to a full chart, allowing encoding and displaying more information.
Analysts can either brush on the context charts or use zoom and pan in the focus view for flexible exploration.
Anomalies are highlighted with red curve segments, which is consistent with the context view (\ano{d1} and \ano{d2} in Fig.~\ref{fig:interface}\cc{d}).
A grey background is further added to make the anomalies more visually apparent.
The anomaly score (Eq.~\ref{eq:score}) is encoded by the colored header bar with a diverging color scheme from blue to red to indicate the severity of the anomaly(\ano{d4} in Fig.~\ref{fig:interface}\cc{d}).
Analysts can create a new event by brushing a new window or modify an existing event by clicking on it (\ano{d2} in Fig.~\ref{fig:interface}\cc{d}, \textbf{R6}).

When there is only one station selected, the focus view adds additional information including the LSTM prediction results and the smoothed errors. Otherwise, this information will be hided.
Prediction results are visualized with a brown curve and the smoothed errors are represented as a centered flow on the top of the chart (\ano{d3} in Fig.~\ref{fig:interface}\cc{d}).
These visualizations enable analysts to visually evaluate the accuracy of the prediction model and reason about why an anomaly is identified, which can greatly facilitate the hyperparameters tuning of LSTM and FindAnomaly block.

\subsubsection{\textbf{Period View}}

Period view (Fig.~\ref{fig:interface}e) is designed for analyzing periodic patterns of a certain type of pollutant. 
Three levels of periodical glyphs in correspondence with the different time granularities (i.e., year, month, and day) are proposed to support multi-scale analysis (\textbf{R4}).
The glyph design uses a circular time axis
Inspired by circular silhouette graph~\cite{aigner2011visualization}, our glyph design uses a circular time axis to emphasize periodicities and the area height encodes the size of values.
We represent each time-sereis as an individual glyph and use small multiples to support side-by-side comparison among multiple time-series.
For example, a month can be divided into many days; each day can be represented with a circular glyph. We use a calendar layout to organize the glyphs (Fig.~\ref{fig:interface}e) so that the daily and weekly patterns will be clearly revealed.
Analysts initially are provided with the year-level glyphs and can move to the fine-grained levels by clicking interactions.




\section{Evaluation}\label{sec:evaluation}

\subsection{ML Pipeline Evaluation}
In this section, we evaluated the overall performance of the proposed machine learning pipeline based on a quantitative study and
demonstrated the usefulness of the system through a set of case studies on the real-world datasets described in Section~\ref{sec:background:data}.
The case studies were completed along with our domain experts; relevant feedback was collected and summarized as a qualitative evaluation.

\textbf{Ground-truth collection}.
We recruited two annotators to manually label the anomalous events regarding \pmtwo~in 5 stations in Hong Kong.
The annotators were asked to search online news thoroughly in SCMP (a popular Hong Kong newspaper founded in 1903) under the column of Hong Kong Air Pollution to document the related events that occured after 2016.
As a complementary, the annotators were further asked to use \name, where the anomaly highlighting function was disabled, to manually identify the time segments where \pmtwo~had extremely abnormal changes in trend.
Finally, 44 events were identified.

\textbf{Performance score computation}.
To calculate the performance score of a given pipeline, the start and end points of the identified anomalies and the ground truth are combined to a sequence of points.
For each interval in this sequence, we check whether there is an anomaly in the ground truth or the identified anomalies.
We create two new sequences, one for ground truth and one for identified anomalies.
For each of the intervals, the two sequences contain a binary value: 0 for no anomaly and 1 for an anomaly. 
The result is two sequences indicating the true anomalous intervals and the identified ones, which then can be used to calculate performance scores such as precision and recall. The length of the intervals is used as the sample weight for the metrics.


\textbf{Result Analysis}. 
Considering there are 44 events in total with each station having 8.8 events on average, we tuned the hyperparameters of both the LSTM and FindAnomaly blocks to produce roughly the same number of anomalies.
Table~\ref{tb:score} summarizes the evaluation results.
The mean precision achieves $83.34\%$, indicating that most of the identified time segments actually exist true anomalous events.
The mean recall is $50.63\%$, suggesting that around half of the labeled events are identified by our pipeline.
In fact, in our application scenario, the experts put more emphasis on precision, as lower false positive rates can facilitate analysts in focusing on evaluating potentially anomalous events.
Hence, we use F\textsubscript{0.5} score as our final performance score.
Both experts thought that the F\textsubscript{0.5} score of $73.8\%$ was acceptable, considering that there is no clear definition of anomalies which might introduce some bias during the label collecting process.
Thus, we further provided a qualitative study in the following to showcase our system. 

\begin{table}[!t]
\centering
\caption{Performance evaluation results of the pipeline.}
\setlength\tabcolsep{4.5pt}
\begin{tabular}{c|cccccc}
Station\# & 1       & 2       & 3       & 4       & 5       & \textbf{mean}    \\ \hline
Precision & 79.6\%  & 80.10\% & 82.30\% & 87.90\% & 86.80\% & \textbf{83.34\%} \\
Recall    & 53.75\% & 51.30\% & 50.40\% & 46.30\% & 51.40\% & \textbf{50.63\%} \\
F\textsubscript{0.5} Score  & 72.62\% & 72.01\% & 73.05\% & 74.51\% & 76.29\% & \textbf{73.8\%} \\
\end{tabular}
\vspace{-10pt}
\label{tb:score}
\end{table}






\subsection{Case Studies}

We qualitatively studied the usability and usefulness of our system through two case studies on the real-world datasets described in Section~\ref{sec:background:data}.
The case studies were completed along with our domain experts.

\subsubsection{Hong Kong Air Quality Analysis}

This case aims to demonstrate the usefulness of the functionalities provided by our system, \name~.

\textbf{Tuning hyperparameters.}
The Hong Kong Air Dataset contains records for around 4 years, which indicates there could be relatively many events related to air quality.
To ensure an efficient analysis (\textbf{R2, R3}), the experts emphasized the importance of reducing the number of anomalies to a reasonable number so that they can focus attention on the most important time segments.
There are several hyperparameters in the pipeline that are highly related to the number of anomalies finally produced. 
The experts iteratively tuned these parameters on the interface, re-ran the experiments, and observed the resulting visualizations (\cc{a} and \cc{c} in Fig.~\ref{fig:interface}).
In the end, a proper setting of hyperparameters was found (Fig.~\ref{fig:interface}\cc{a}), and the corresponding pipeline was applied to 16 stations in Hong Kong (\textbf{R1, R2}).
Note that the average event number detected for every time-series is $11.5$, which is slightly higher than the averaged event number (8.8) collected by the annotators. 

\textbf{Exploring air quality dynamics.}
The experts first examined a roadside station, ``Causeway Bay'', seated nearby the city center (\ano{b1} of Fig.~\ref{fig:interface}\cc{b}).
From Fig.~\ref{fig:interface}\cc{c}, they could browse the overall dynamics of every pollutant and weather variable, as well as how the anomalous events were distributed over time (\textbf{R3, R4, R5}).
Two time segments (\ano{c1} and \ano{c2}) initially caught their attention for the interesting co-occurrence phenomenon.
In the first time segment \ano{c1}, there were anomalous events detected regarding pressure, wind speed, \pmtwo~and \pmten.
The experts inferred that a strong airflow transported pollutants to Hong Kong, sending the pollutants away and returning pollution levels back to normal in a very short time.
Another interesting finding is time segment \ano{c2}, where two high co-peaks of \pmtwo, and \pmten~appeared around Jan. 12 and a relatively high co-peaks regarding \pmten, \sotwo, and CO were identified few days ago.
In both time segments, no obviously unusual behaviors were detected regarding weather.


\textbf{Assessing and editing anomalous events.}
The experts turned to the focus view for further examination (\textbf{R4}).
Take \pmtwo~as an example. The peak \ano{d1} is the detected anomalous event, where the predicted value deviates much from the true value, thereby causing high prediction errors (\ano{d3}).
The experts then found the other relatively high peak (\ano{d2}) which was not identified as an anomaly by the system.
The period view clearly showed the two peaks appeared on neighboring Saturdays (\ano{e1} in Fig.~\ref{fig:interface}\cc{e}).
Considering there are co-peaks regarding \pmten, \sotwo, and CO on the first Saturday.
The experts thereby suspected that the first Saturday could be the starting point of the event which transpired the next Saturday.
Hence, they manually created an event on \pmtwo with severity level 2 and made a comment accordingly (\ano{d2}, \textbf{R6}).
Based on this, they further conjectured that the following few days would be windless, leading to the higher peak of pollutions on Jan 13.
This was verified by the examination in the focus view of wind speed (\ano{c3}), where the wind speed of the day before the second Saturday dropped almost to zero.
The above finding was further demonstrated by the air quality news report\footnote{Windless conditions worsen air quality. \url{https://tinyurl.com/ycrqtzop}}, where the reported finding was consistent with ours.

\begin{figure}[!tbp]
	\centering
	\vspace{-12pt}
	\includegraphics[width=1\linewidth]{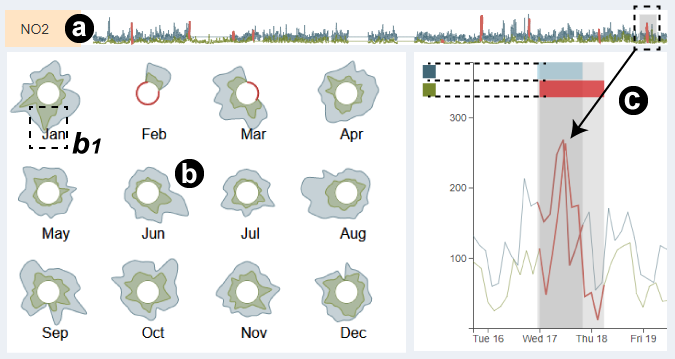}
	\vspace{-18pt}
	\caption{\cc{a}\cc{b} Much more \notwo~is detected at the roadside station Causeway Bay (blue) compared with the general station Tung Chung (green). \cc{c} The severity level of the anomalous event regarding Tung Chung is higher than the one regarding Causeway Bay.}
	\vspace{-12pt}
	\label{fig:hk_compare}
\end{figure}

\textbf{Comparing different stations.}
The experts were curious about the difference between stations (\textbf{R5}).
Therefore, they chose a non-roadside station that is far away from the city center for a comparative analysis (Tung Chung, \ano{b2} in Fig.~\ref{fig:interface}\cc{b}).
From the context view, they found that most of the pollutants shared similar variation patterns except for \notwo~(Fig.~\ref{fig:hk_compare}\cc{a}).
The pattern could be revealed more clearly in the period view (Fig.~\ref{fig:hk_compare}\cc{b} shows the month-level glyphs of the year 2018).
This could be explained by the fact that the Causeway Bay station (blue) is a roadside station which is near by heavy traffic, which would produce lots of \notwo.
The experts further observed that the green and the blue areas have co-peaks in the middle of January (\ano{b1}).
In the focus view (Fig.~\ref{fig:hk_compare}\cc{c}), the experts found that the anomalous event coming from the green station (Tung Chung) has higher severity level, as the green station reached a new peak with the same level as the blue station that usually had higher \notwo.
This suggests that air quality hit very unhealthy level in many regions of Hong Kong in this time period and \notwo~was one of the major pollutant; there was a news report fully supporting this conjecture\footnote{Air pollution blanketed multiple areas. \url{https://tinyurl.com/yxhxa22a}}.

\subsubsection{Beijing and Shenzhen Air Quality}

This case focuses on qualitatively evaluating the general applicability and adaptability of the pipeline.

The experts expressed strong interest in exploring air quality in different cities, as conditions could be diverge significantly in different cities.
Hence, they further explored two metropolitan cities in China, Beijing and Shenzhen, using the same pipeline as the previous case study to detect anomalous events.

\begin{figure}[!tbp]
	\centering
	\vspace{-12pt}
	\includegraphics[width=1\linewidth]{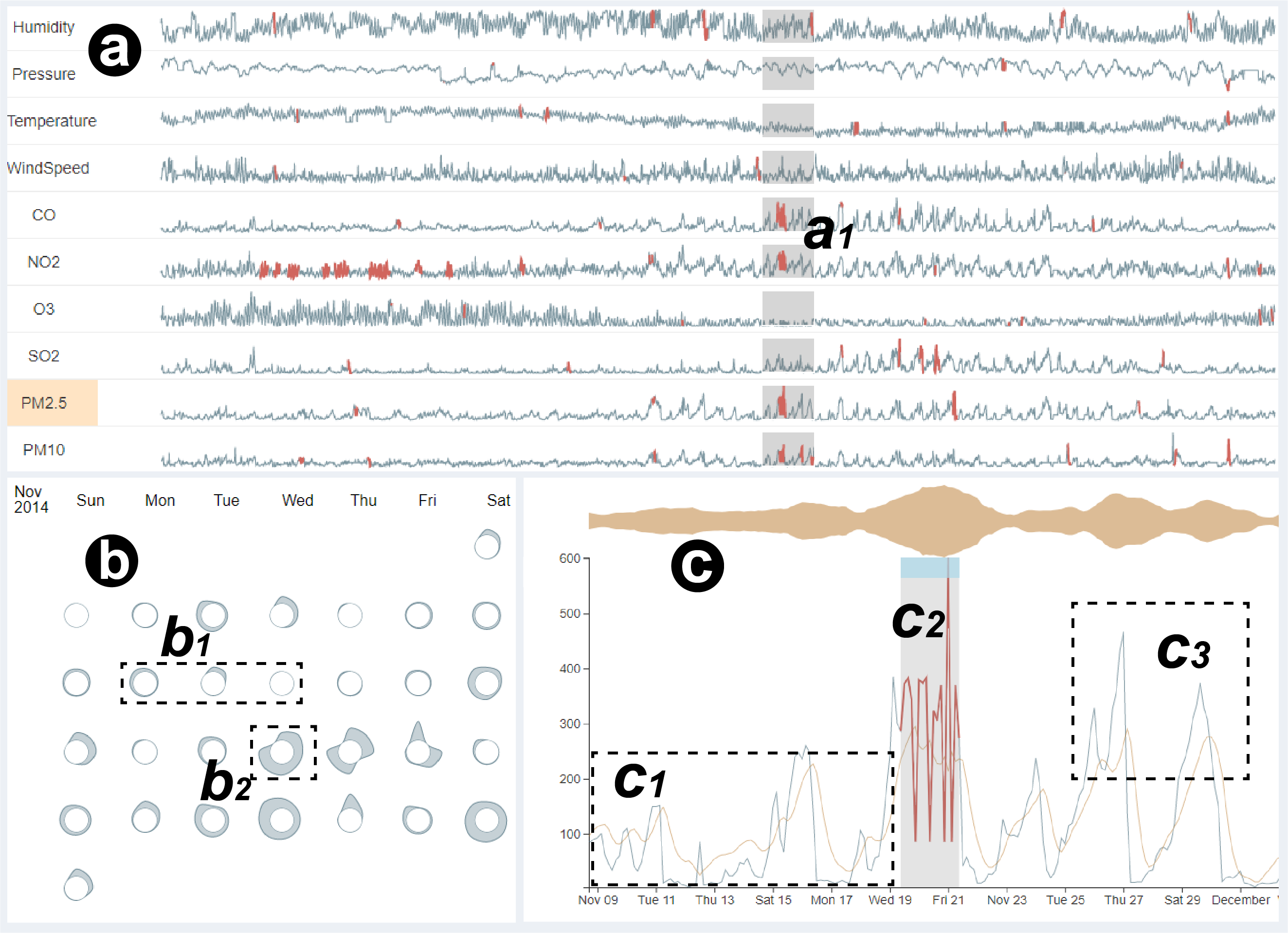}
	\vspace{-18pt}
	\caption{\cc{a} The air quality in Beijing is much more complicated for the frequently violent fluctuations in various pollutants in Beijing. \cc{b}\cc{c} The air quality was at a very low level of pollution in the days around Nov. 10 (\ano{b1}, \ano{c1}) due to the holding of APEC China 2014 in Beijing. The first yellow pollution alert was issued one week following the summit (\ano{b2}, \ano{c2}).}
	\vspace{-14pt}
	\label{fig:bj}
\end{figure}

\textbf{General applicability}. Fig~\ref{fig:bj} shows the information of a station nearby the city center of Beijing.
The experts' initial impression of the dataset was that the air quality in Beijing was much more complicated due to the observation of frequently violent fluctuations in various pollutants (Fig.~\ref{fig:bj}\cc{a}).
Their attention was attracted by one interesting co-occurrence pattern (\ano{a1}), where the anomalous events regarding four pollutants were detected with no abnormal changes in weather variables.
The three pollutants including CO, \pmtwo, and \pmten~are kept in a very low level for several weeks before the detected time segment, which could be further verified from the focus view (\ano{c1} and \ano{c2}) and period view (the days before \ano{b2}).
The experts suspected that this phenomenon could be triggered by some special events.
This is verified by the news\footnote{Farewell to 'APEC blue'. \url{https://tinyurl.com/y282arxo}} that Beijing was holding APEC China 2014 from Nov. 10 to Nov. 12 (\ano{b1}) and before that time the government requested a reduction in the emission of pollutants (aka APEC blue\footnote{APEC blue. \url{https://en.wikipedia.org/wiki/APEC_blue}}). 

\begin{figure}[!t]
	\centering
	\includegraphics[width=1\linewidth]{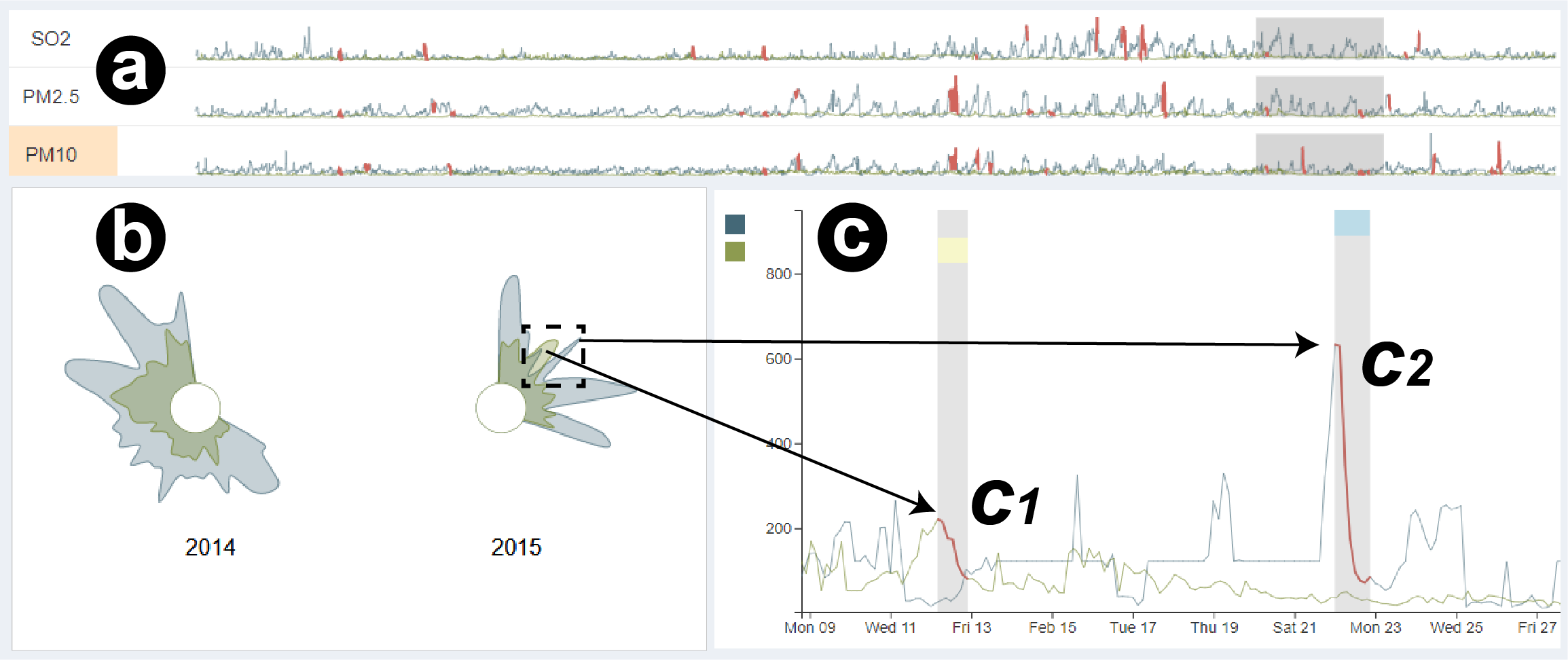}
	\vspace{-16pt}
	\caption{\cc{a}\cc{b} Three pollutants in Beijing (blue) are significantly higher than in Shenzhen (green). \cc{c} The pipeline learns different standards in different cities. The anomalous event regarding \pmtwo~detected in Shenzhen (\ano{c1}) has higher severity though the actual values are much lower than the one in Beijing (\ano{c2}).}
	\vspace{-12pt}
	\label{fig:bjsz}
\end{figure}

\textbf{Intelligence and Adaptability}. Furthermore, from the focus view of \pmtwo~(Fig.~\ref{fig:bj}\cc{c}), the experts realized the intelligence of the anomaly detection pipeline.
First, the pipeline detected the first significant peak after a long ``peaceful'' time, and intelligently classified the subsequent close peaks as normal cases because they are not unexpected (\ano{c3}).
This could be explained by that the temporal context was leveraged by the LSTM model to detect anomalies~\cite{ahmad2017unsupervised}.
Second, the pipeline learned different standards to adapt itself to different environments.
This was verified when experts selected a station from Shenzhen (in South China) to compare with the current station from Beijing (North China).
As shown in \cc{a} and \cc{b} of Fig.~\ref{fig:bjsz}, the three pollutants including \sotwo, \pmtwo, and \pmten~in Beijing (blue) were significantly higher than in Shenzhen (green).
The experts confirmed that the pipeline had a good adaptivity as the anomaly highlighted in the green line (\ano{c1}) had an even higher anomaly score despite that the actual values were much lower than the anomaly in the blue line (\ano{c2}).

\subsection{User Feedback}

During the process of these case studies, we have collected a wealth of valuable feedback from the experts which could be summarized in the following three aspects.
(1) \textbf{Machine learning pipeline:}
The experts were satisfied with the design of our unified end-to-end tunable ML pipeline for air quality data analysis.
EA commented that ``The pipeline relieves me from the pain of handling data quality issues.  The hyperparameters are organized with a clear structure, so that I can easily know what to change through the visual interface to start a new experiment.''
EB especially praised the flexibility of the pipeline and commended, ``In the long run, we can integrate more blocks to support more tasks.''
(2) \textbf{Visualization and interaction:}
Feedback showed that most of the visualization designs met our requirements.
For example, the experts felt the time-series context view together with the focus view indeed saved them time finding the pollutants and time segments of interest for detailed analysis.
EA also commented, ``The visualizations of prediction results, errors, and anomaly score increase my confidence on the ML pipeline and also show me a direction to debug and optimize the pipeline.''
Both experts gave high praise to the period view.
EA commented, ``The glyph design is intuitive and very helpful in observing periodical patterns. The small multiples layout allow me to effectively compare the difference among days, months, and years.''
Nevertheless, EB also pointed out a feature that the current system cannot support: ``I have to go back and forth to compare the same month such as July in different years.''
The interactions were also highly praised by the experts. ``The system integrates many animation, navigation and interaction techniques, allowing me to easily play with the air quality data.''
(3) \textbf{Usability:}
The experts were confident that the system would offer tremendous aid in analyzing air quality data and in the writing of a summary report of air quality focused on anomalies.
They emphasized that previously only the number of exceedances of AQO (Air Quality Objectives) limit value and some statistical values were reported for a given pollutant concentration.
``With this tool, more types of anomalies can be identified and analyzed in further detail.''


\section{Discussion and Future Work}

The effectiveness of our system is demonstrated in the evaluation section.
Nevertheless, there is still space for improvement and many potential future directions worthy of further investigation.

\textbf{Adaptivity of the Pipeline.}
The current system applies one pipeline with the same hyperparameter setting over different stations.
Although the current results are good due to the excellent adaptivity of LSTMs and the error threshold learning algorithm, performance can be further improved.
For example, we can employ Bayesian hyperparameter optimization methods~\cite{snoek2012practical, gustafson2018bayesian} on the learning blocks of each individual pipeline at every station to find better hyperparameter configurations. 
In addition, our system allows analysts to interactively edit the anomalies as well as add tags or comments.
We can investigate how to leverage such feedback to fine tune the pipeline accordingly.

\textbf{Scalability of the Visualization.}
When too many stations (more than four) are selected at the same time for comparative analysis, a serious visual clutter problem would emerge, due to the limited screen space and the limited number of categorical colors that user can effectively differentiate~\cite{munzner2014visualization}.
We believe our designs work for most cases as analysts typically only need to compare two stations. In the meantime, we also plan to investigate more design choices to better support comparative analysis among many stations.

\textbf{Intepretibility of the Anomalies.}
Though we have provided a couple of visualizations to explain the anomalies - including the visualizations of anomaly scores, smoothed errors, and prediction results - analysts still lack a direct platform to reason with why the anomalous event occurred.
There are two directions for potential future research which may improve the interpretability of anomalies.
The first is revealing more detailed information of the algorithms.
For example, we can use an occlusion-based method to evaluate which parts of an input sequence contribute most to the prediction result~\cite{li2016understanding}.
The second is integrating more information from different sources, such as local news and regional weather photos.

\textbf{Capability of the Real-time Analysis.}
The current system is mainly used for offline analysis of historical data.
To support real-time analysis, we can run the ML pipeline at every station every few days to keep the models up-to-date and learning the latest dynamic thresholds.
These updated models and thresholds could then be used to predict anomalies as new data arrives over the next several days.
For visualization, many of the features can be kept the same, but it is also possible to integrate several stream data visualization techniques~\cite{wu2018streamexplorer} to better highlight newly-arrived data.

\section{Conclusion}\label{sec:conclusion}

In this paper, we propose a novel visual analytics system, \name, to disclose the rich dynamics of air quality data and support efficient exploration and examination of anomalous air quality events.
The system is built based on a highly flexible and scalable machine learning pipeline, which can handle multiple data challenges introduced by real-world air quality datasets, such as missing values and inconsistent data formats.
Specifically, the pipeline integrates an efficient LSTM-based unsupervised anomaly detection method that works without the use of labeled data and is able to diversified anomalous events of high quality.
The effectiveness and usefulness of the system is evaluated through a quantitative study and two case studies on real-world datasets.
The possible feedback from the experts further confirms the strengths of our system.

\section*{Acknowledgment}
This research is supported in part by the Theme-based Research Scheme of the Research Grants Council of Hong Kong,  under Grant No. T41-709/17-N.


\bibliographystyle{IEEEtran}
\bibliography{reference}

\end{document}